\documentclass[final]{siamltex} 
\usepackage{float}
\usepackage{amssymb} 
\usepackage{amsmath}
\usepackage{graphicx}
\usepackage[outdir=./]{epstopdf}
\usepackage{tikz} 
\usetikzlibrary{fit,positioning}
\usepackage{subcaption}
\usepackage{algorithmicx}
\usepackage{algorithm}
\usepackage{algpseudocode}
\usepackage{placeins}
\usepackage{bm}
\usepackage{amsfonts}
\usepackage{amsopn}
\usepackage[utf8]{inputenc}
\usepackage[framed,numbered,autolinebreaks,useliterate]{mcode}
\usepackage{caption}
\usepackage[colorlinks=true]{hyperref}
\usepackage{comment}
\usepackage{url}
\makeatletter
\g@addto@macro{\UrlBreaks}{\UrlOrds}
\makeatother

\usepackage{tikz-network}

\usepackage{draftwatermark}

\usepackage{listings}
\tikzstyle{int}=[draw, fill=white!8, minimum size=2em]
\tikzstyle{init} = [pin edge={to-,thin,black}]

\usepackage{tikz}
\usetikzlibrary{matrix}
\usetikzlibrary{calc}
\usetikzlibrary{arrows}
\usepackage{relsize}
\usepackage[export]{adjustbox}
\usepackage{array,longtable,multirow}

\usepackage{xcolor,colortbl}

\definecolor{Gray}{gray}{0.95}
\newcolumntype{a}{>{\columncolor{Gray}}c}

\algblock{ParFor}{EndParFor}
\algnewcommand\algorithmicparfor{\textbf{parfor}}
\algnewcommand\algorithmicpardo{\textbf{do}}
\algnewcommand\algorithmicendparfor{\textbf{end\ parfor}}
\algrenewtext{ParFor}[1]{\algorithmicparfor\ #1\ \algorithmicpardo}
\algrenewtext{EndParFor}{\algorithmicendparfor}

\newtheorem{remark}[theorem]{Remark}

\title{A Multilayer Network Model implementation for Covid-19}

\author{Juan G. Calvo\thanks{Universidad de Costa Rica, Centro de Investigaci\'on en Matem\'atica Pura y Aplicada - Escuela de Matem\'atica, San Jos\'e, Costa Rica ({juan.calvo@ucr.ac.cr}, {fabio.sanchez@ucr.ac.cr}, {luisalberto.barboza@ucr.ac.cr}).}
\and Fabio Sanchez\footnotemark[1]
\and Luis A. Barboza\footnotemark[1]
\and Yury E. García\thanks{Universidad de Costa Rica, Centro de Investigaci\'on en Matem\'atica Pura y Aplicada, San Jos\'e, Costa Rica ({paola.vasquez@ucr.ac.cr}, {yury.garciapuerta@ucr.ac.cr}).}
\and Paola V\'asquez\footnotemark[2]}

\begin{document}

\SetWatermarkText{DRAFT}

\maketitle

\begin{abstract}
We present a numerical implementation for a multilayer network used to model the transmission of Covid-19 or other diseases with a similar transmission mechanism. The model incorporates different contact types between individuals ({\it household}, {\it social contacts}, and {\it strangers}), which allows flexibility compared to standard SIR type models. The algorithm described in this paper is a simplification of the model used to give public health authorities an additional tool for the decision-making process in Costa Rica, by simulating extensive possible scenarios and projections.
\end{abstract}



\section{Introduction}
Throughout the Covid-19 pandemic, mathematical and statistical models have been used as a way to forecast or analyze the behavior of the disease in various parts of the world \cite{COVID19F37,adiga2020mathematical,torrealba2020modeling,kucharski2020early,London}. Different mathematical tools have been implemented to understand the disease transmission and to plan for strategic sanitary measures to minimize the burden on the public health system \cite{COVID19F37,London,star2010role}. Deterministic models have been widely used to model infectious diseases in the past \cite{perelson2013modeling, dietz1985mathematical, etbaigha2018seir,hethcote2000mathematics,choisy2007mathematical,chen2012modeling}. However, due to human mobility, unpredictable human behavior and other factors, adaptable models are better suited to evolve with the ever-changing behavior of the population during the pandemic as sanitary measures are implemented in a particular region of interest.

In this paper, we consider an adaptive multilayer network, capable of managing large populations efficiently; see \cite{multilayer1,multilayer2}. A probability of infection is computed on a daily basis for each node independently, permitting a parallel implementation for real-life applications. We count daily interactions per individual and determine if the virus is transmitted depending on the attributes of each interaction. 

One of the main advantages of a network model is its flexibility when taking into account individual and social behavior, that allows us to mimic sanitary measures and evaluate possible impacts on the population. As the pandemic progresses, it is possible to incorporate early or ongoing behavioral changes such as reductions in community interactions, mask use, mobility restrictions, network contacts, and lockdowns. We include a Matlab app with the algorithm implementation that can be found in \cite{app}, where several attributes and parameters can be modified\footnote{A full open code will be available as well.}.

The model we present in this paper is a simplification of the model that has been developed and taken into consideration by the health authorities in Costa Rica. It includes the entire population (approximately five million individuals) and the 81\footnote{There are 82 municipalities in the country. However, the 82nd was established in 2018 and there is no historical census data in order to model mobility.} municipalities of the country. The model also considers different scenarios with gradual re-openings, mobility restrictions, different percentages for mask use, as well as short and long-term projections of the disease. Results were delivered to the public health authorities as part of the decision-making process \cite{PAHO3,PAHO4,PAHO5,PAHO6}. The model takes into account local data, public health policies pertinent to the country and more specific details than the model described in this work.

The rest of the paper is organized as follows. In Section \ref{sec:prelim}, we describe the algorithm. A pseudo-code is presented in Section \ref{sec:impl} with a discussion of the implementation. In Section \ref{sec:example}, we present two toy examples, in order to show the behaviour of the model. Finally, we include some conclusions and future work in Section \ref{sec:conc}.

\section{Preliminaries} \label{sec:prelim}
Consider a constant population that lives in a fixed number of counties. Mathematically, we consider a graph (a set of nodes that are connected by edges) where each node represents one individual. Interactions are classified into three \textit{layers}, which represent a different type of contact with particular characteristics that account for the virus transmission. These layers are:
(1) a {\it household network} (people that live in the same house), (2) a {\it social network} (known contacts such as friends and colleagues), and (3) a {\it sporadic network} (strangers that you may encounter in short periods of time when you visit random locations). 

An edge between two nodes represents a feasible interaction between two individuals at a particular time, with specific attributes for its interaction depending on the layer. These layers are randomly generated. Layers 1 and 2 are fixed for each simulation and layer 3 can change for each time step, since usually a node has no control on sporadic encounters.

For the first layer, it is assumed that we know the total number of households per county, and the average number of individuals per household per county; the information for Costa Rica is available at \cite{INEC}. We then group all the individuals into families. The number of members per household is chosen by using a Poisson distribution with mean equal to the average number of individuals per household per county. We assume total connectivity for each family and no edges between different families. It is possible to randomly assign an age group to each family member if data is available, in order to include differences in interactions accordingly; see further details in Section \ref{sub:households}.

For the second and third layers, choosing the contacts of a node requires some assumptions. It is natural to expect a different number of interactions depending on age, location, density, socioeconomic factors and more. In our implementation, for every node we need to: (1) define the degree of the node (number of contacts) per layer, given by a uniform distribution on an interval with endpoints equal to the minimum and maximum number of allowed contacts per county per layer, and (2) choose randomly its contacts from different counties based on a given connectivity distribution between counties; see Section \ref{sub:updateNetwork}.

Each node on the graph includes several attributes. We classify them in two classes: (1) fixed parameters, such as county, household members, age group, degree and graph connectivity for layer 2, and (2) variable attributes such as connectivity for layer 3, epidemiological state, number of days at current epidemiological state, number of interactions with social contacts per day, mask use and self-care behaviour. The former are defined at the beginning of each simulation, and the latter can take random values every time step accordingly.

\begin{figure}[ht!]
\centering
\begin{tikzpicture}[node distance=2.5cm,auto,>=latex']
    \node [int, fill=blue!5!white] (S) {$S$};
    \node [int, fill=blue!5!white] (E) [right of= S] {$E$};
    \node [int, fill=blue!5!white] (O) [right of= E] {$O$};
    \node [int, fill=blue!5!white] (H) [right of= O] {$H$};
    \node [int, fill=blue!5!white] (R) [right of= H] {$R$};
    \node [int, fill=blue!5!white] (U) [below of= O, yshift=0.5cm] {$U$};
    \node [int, fill=blue!5!white] (D) [below of= H, yshift=1.1cm] {$D$};

    \path[->] (S) edge node {} (E);
    \path[->] (E) edge [above left] node {} (O);
    \path[->] (O) edge node {} (H);
    \path[->] (U) edge [below right] node {} (H);
    \path[->] (H) edge node {} (R);
    \path[->] (H) edge node {} (D);
    
    \draw[->] (E) |- node[above right=0.5pt,pos=0.7] {} (U);
    \draw[->] (U) -| node[pos=0.25] {}(R);
    \draw[->] (O) -- +(0,1.2) -| node[pos=0.25] {} (R);
\end{tikzpicture}
\caption{Model compartment states and transitions.} 
\label{fig:model1}
\end{figure}
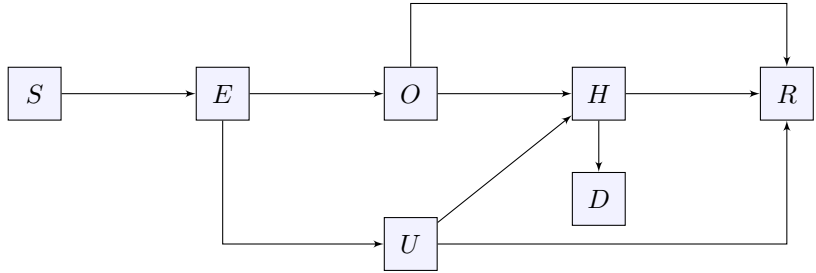

For simplicity, the propagation in our model is based on an SIR type model with seven compartment states defined by the following epidemiological variables: susceptible ($S$), exposed ($E$), diagnosed or observed ($O$), undiagnosed or not observed ($U$), hospitalized ($H$), recovered ($R$), and dead ($D$) states; see Figure \ref{fig:model1}. Modifications with additional states and transitions are straightforward to include. 

\begin{figure}[ht!]
\begin{subfigure}{.5\textwidth}
\centering
  \begin{tikzpicture}[scale=.5,multilayer=3d]
    \SetLayerDistance{-4}
    \Vertices[NoLabel,opacity=0]{data/00_vertices.csv}
    \begin{Layer}[layer=3]
      \draw[orange,very thick,fill=white,fill opacity=.7] (1.5,-.5) rectangle (6.5,4.5);
      \node at (1.5,-.5)[below right,black]{\Huge Sporadic};
    \end{Layer}
    \Edges[NotInBG,layer={3,3}]{data/00_edges.csv}
    \Edges[NotInBG,layer={2,3}]{data/00_edges.csv}
    \Vertices[layer=3]{data/00_vertices.csv}
    \begin{Layer}[layer=2]
      \draw[orange,very thick,fill=white,fill opacity=.7] (1.5,-.5) rectangle (6.5,4.5);
      \node at (1.5,-.5)[below right,black]{\Huge Social};
    \end{Layer}
    \Edges[NotInBG,layer={2,2}]{data/00_edges.csv}
    \Edges[NotInBG,layer={1,2}]{data/00_edges.csv}
    \Vertices[layer=2]{data/00_vertices.csv}
    \begin{Layer}[layer=1]
      \draw[orange,very thick,fill=white,fill opacity=.7] (1.5,-.5) rectangle (6.5,4.5);
      \node at (1.5,-.5)[below right,black]{\Huge Household};
    \end{Layer}
    \Edges[NotInBG,layer={1,1}]{data/00_edges.csv}
    \Vertices[layer=1]{data/00_vertices.csv}
  \end{tikzpicture}
  \captionof{figure}{An infectious node and its network.}\label{fig:network1}
\end{subfigure}%
\begin{subfigure}{.5\textwidth}
\centering
  \begin{tikzpicture}[scale=.5,multilayer=3d]
    \SetLayerDistance{-4}
    \Vertices[NoLabel,opacity=0]{data/01_vertices.csv}
    \begin{Layer}[layer=3]
      \draw[orange,very thick,fill=white,fill opacity=.7] (1.5,-.5) rectangle (6.5,4.5);
      \node at (1.5,-.5)[below right,black]{\Huge Sporadic};
    \end{Layer}
    \Edges[NotInBG,layer={3,3}]{data/01_edges.csv}
    \Edges[NotInBG,layer={2,3}]{data/01_edges.csv}
    \Vertices[layer=3]{data/01_vertices.csv}
    \begin{Layer}[layer=2]
      \draw[orange,very thick,fill=white,fill opacity=.7] (1.5,-.5) rectangle (6.5,4.5);
      \node at (1.5,-.5)[below right,black]{\Huge Social};
    \end{Layer}
    \Edges[NotInBG,layer={2,2}]{data/01_edges.csv}
    \Edges[NotInBG,layer={1,2}]{data/01_edges.csv}
    \Vertices[layer=2]{data/01_vertices.csv}
    \begin{Layer}[layer=1]
      \draw[orange,very thick,fill=white,fill opacity=.7] (1.5,-.5) rectangle (6.5,4.5);
      \node at (1.5,-.5)[below right,black]{\Huge Household};
    \end{Layer}
    \Edges[NotInBG,layer={1,1}]{data/01_edges.csv}
    \Vertices[layer=1]{data/01_vertices.csv}
  \end{tikzpicture}
  \captionof{figure}{Two new infected nodes.}\label{fig:network2}
\end{subfigure}
\begin{subfigure}{.5\textwidth}
\centering
  \begin{tikzpicture}[scale=.5,multilayer=3d]
    \SetLayerDistance{-4}
    \Vertices[NoLabel,opacity=0]{data/02_vertices.csv}
    \begin{Layer}[layer=3]
      \draw[orange,very thick,fill=white,fill opacity=.7] (-.5,-.5) rectangle (6.5,4.5);
      \node at (-.5,-.5)[below right,black]{\Huge Sporadic};
    \end{Layer}
    \Edges[NotInBG,layer={3,3}]{data/02_edges.csv}
    \Edges[NotInBG,layer={2,3}]{data/02_edges.csv}
    \Vertices[layer=3]{data/02_vertices.csv}
    \begin{Layer}[layer=2]
      \draw[orange,very thick,fill=white,fill opacity=.7] (-.5,-.5) rectangle (6.5,4.5);
      \node at (-.5,-.5)[below right,black]{\Huge Social};
    \end{Layer}
    \Edges[NotInBG,layer={2,2}]{data/02_edges.csv}
    \Edges[NotInBG,layer={1,2}]{data/02_edges.csv}
    \Vertices[layer=2]{data/02_vertices.csv}
    \begin{Layer}[layer=1]
      \draw[orange,very thick,fill=white,fill opacity=.7] (-.5,-.5) rectangle (6.5,4.5);
      \node at (-.5,-.5)[below right,black]{\Huge Household};
    \end{Layer}
    \Edges[NotInBG,layer={1,1}]{data/02_edges.csv}
    \Vertices[layer=1]{data/02_vertices.csv}
  \end{tikzpicture}
  \captionof{figure}{Augmented network.}\label{fig:network3}
\end{subfigure}%
\begin{subfigure}{.5\textwidth}
  \centering
  \begin{tikzpicture}[scale=.5,multilayer=3d]
    \SetLayerDistance{-4}
    \Vertices[NoLabel,opacity=0]{data/03_vertices.csv}
    \begin{Layer}[layer=3]
      \draw[orange,very thick,fill=white,fill opacity=.7] (-.5,-.5) rectangle (6.5,4.5);
      \node at (-.5,-.5)[below right,black]{\Huge Sporadic};
    \end{Layer}
    \Edges[NotInBG,layer={3,3}]{data/03_edges.csv}
    \Edges[NotInBG,layer={2,3}]{data/03_edges.csv}
    \Vertices[layer=3]{data/03_vertices.csv}
    \begin{Layer}[layer=2]
      \draw[orange,very thick,fill=white,fill opacity=.7] (-.5,-.5) rectangle (6.5,4.5);
      \node at (-.5,-.5)[below right,black]{\Huge Social};
    \end{Layer}
    \Edges[NotInBG,layer={2,2}]{data/03_edges.csv}
    \Edges[NotInBG,layer={1,2}]{data/03_edges.csv}
    \Vertices[layer=2]{data/03_vertices.csv}
    \begin{Layer}[layer=1]
      \draw[orange,very thick,fill=white,fill opacity=.7] (-.5,-.5) rectangle (6.5,4.5);
      \node at (-.5,-.5)[below right,black]{\Huge Household};
    \end{Layer}
    \Edges[NotInBG,layer={1,1}]{data/03_edges.csv}
    \Vertices[layer=1]{data/03_vertices.csv}
  \end{tikzpicture}
  \captionof{figure}{Four new infected nodes.}\label{fig:network4}
\end{subfigure}
\caption{Modeling the transmission of the virus. The network adapts on a daily basis depending on new exposed nodes, imitating the evolution of new cases. \label{fig:network}}
\end{figure}
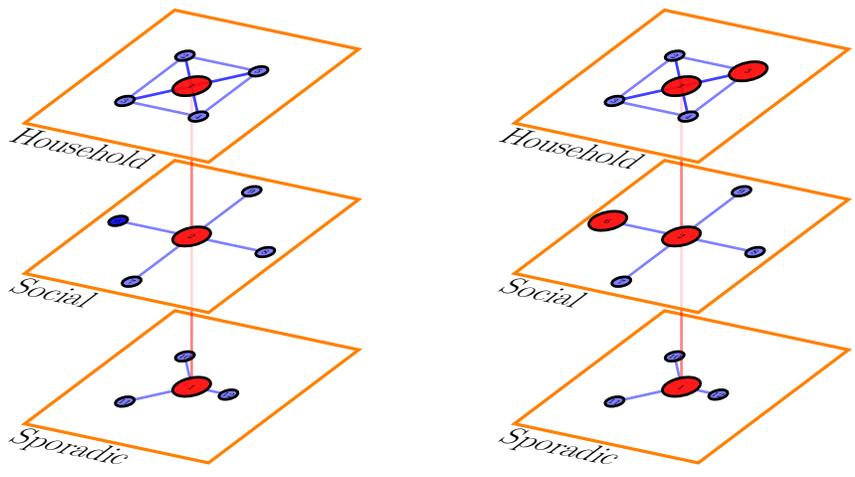
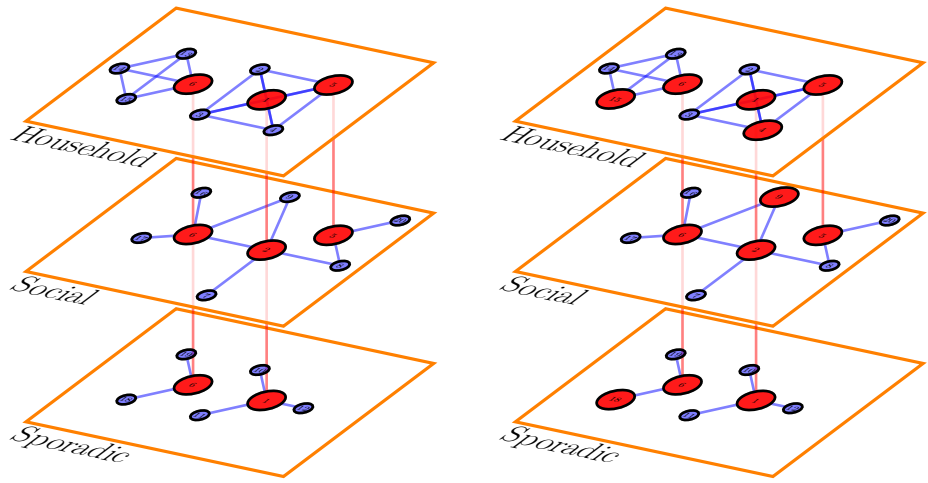

Once the multilayer network is defined, we model the transmission of the virus by computing probabilities of infection $p_i$ for interactions between infectious and susceptible nodes; see Section \ref{sub:probInf}. For a given initial set of exposed nodes, we consider their contacts via the different layers; see Figure \ref{fig:network1}. Depending on $p_i$, adjacent nodes of infectious nodes can become exposed; see Figure \ref{fig:network2}. We then increase the network by including new exposed nodes and their contacts as shown in Figure \ref{fig:network3}, continuing the process on a daily basis; see Figure \ref{fig:network4}.

\begin{remark} \label{rm:1} {\rm
The degree of each node in layer 2 is fixed for each simulation. Nevertheless, on each time step a subset of edges is selected because only a small number of daily contacts is expected to occur. As mentioned before, layer 3 changes daily.
}
\end{remark}

\begin{remark} \label{rem:limitations} {\rm
If the population has a large number of individuals, it is not recommended to create the whole graph initially due to execution times. Instead, at every time step we add new nodes and their edges on the graph in order to achieve faster running times; see Figure \ref{fig:network} and Section \ref{sub:updateNetwork} for further details.
}
\end{remark}

\section{Implementation} \label{sec:impl}
In this section we present a pseudo-code for the modeling of Covid-19 as described in Section \ref{sec:prelim}; see Algorithm \ref{alg:graph}. We describe each step in the following sections in a general and simplified setting.

\begin{algorithm}
\caption{Multilayer network model} \label{alg:graph}
\begin{algorithmic}[1] 
\Procedure {networkModel}{}
\State Create household network. \Comment{See Section \ref{sub:households}}
\State Define initial conditions.\Comment{See Section \ref{sub:initCond}} 
 \For{day = $1,2,3\ldots$}
 \For{all new exposed nodes $E$} \Comment{See Section \ref{sub:updateNetwork}}
  \If{family has not been added}
\State Read connections in the family layer.
\State Store edge connectivity.
\EndIf
\If{node requires more contacts in layer 2}
\State Create list of eligible contacts.
\State Choose random nodes for layer 2.
\State Store edge connectivity for layer 2.
\EndIf
 \EndFor
 \For{all infectious nodes}
 \State Create list of eligible contacts.
 \State Choose random nodes for layer 3.
 \EndFor
 \For{all susceptible nodes $S$} \Comment{See Section \ref{sub:probInf}}
\State Compute probability of infection.
\State Determine if current node has been infected. 
 \EndFor
\State Determine transitions between states.\Comment{See Section \ref{sub:transitions}}
\State Store results for current day.
\EndFor
 \EndProcedure
\end{algorithmic}
\end{algorithm}

\subsection{Creation of household network} \label{sub:households}
In this implementation we assume that the following data is available: total number of households per county, average number of individuals per household per county and age groups; see a toy example in Table \ref{table:familyInfo}. We then create all the households and assign their inhabitants by using a Poisson distribution with mean equal to the average number of individuals per household per county; see Figure \ref{fig:families}. 

Each individual is labeled with an ID from $1$ to $N$, where $N$ is the size of the population. We create the arrays \texttt{IDtoCounty}, \texttt{IDtoFamily} and \texttt{IDtoAge} that return the county, family and age group of a given ID, and the cells \texttt{familyToIDs} and \texttt{countyToIDs} that return the IDs for a given family or county, respectively. These arrays allows us to directly access required information to construct the graph.

\begin{minipage}[t]{\linewidth}%
\centering
\begin{tikzpicture}[multilayer]
  \Vertex[x=0,y=0.5,IdAsLabel,layer=1,color=blue,opacity =.3]{1}
  \Vertex[x=2,y=0.5,IdAsLabel,layer=1,color=red,opacity =.3]{2}
  \Vertex[x=0,y=2.5,IdAsLabel,layer=1]{3}
  \Vertex[x=2,y=2.5,IdAsLabel,layer=1]{4}
  \Vertex[x=5,y=0,IdAsLabel,layer=1]{5}
  \Vertex[x=7,y=0,IdAsLabel,layer=1]{6}
  \Vertex[x=4,y=2,IdAsLabel,layer=1]{7}
  \Vertex[x=6,y=3,IdAsLabel,layer=1,color=blue,opacity =.3]{8}
  \Vertex[x=8,y=2,IdAsLabel,layer=1,color=blue,opacity =.3]{9}
  \Vertex[x=10,y=1.5,IdAsLabel,layer=1,color=blue,opacity =.3]{10}
  \Edge(1)(2)
  \Edge(1)(3)
  \Edge(1)(4)
  \Edge(2)(3)
  \Edge(2)(4)
  \Edge(3)(4)
  \Edge(5)(6)
  \Edge(5)(7)
  \Edge(5)(8)
  \Edge(5)(9)
  \Edge(6)(7)
  \Edge(6)(8)
  \Edge(6)(9)
  \Edge(7)(8)
  \Edge(7)(9)
  \Edge(8)(9)
\end{tikzpicture}

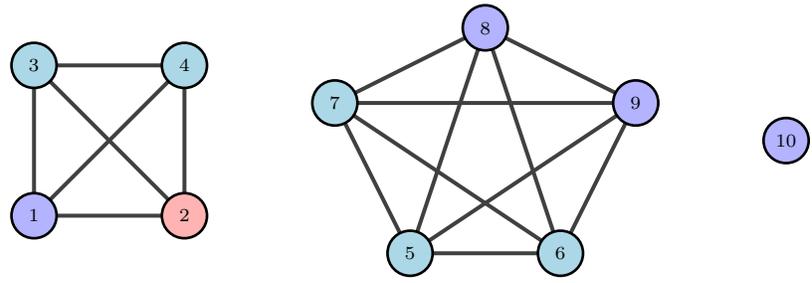
\captionof{figure}{Three families with 4, 5 and 1 members. A fully connected graph for \\ each component is assumed. Colors represent different age groups. \label{fig:families}}
\end{minipage}

\subsection{Initial conditions} \label{sub:initCond}
For simplicity, we assume an initial graph given by all the exposed individuals at time zero and their contact networks; see Figure \ref{fig:network1} for the case of one initial exposed node. If data is available, we can include chains of transmission for a fixed time as an initial condition, by creating edges among identified clusters. Before community transmission in Costa Rica, the public health authorities were able to keep track of chains of transmission for approximately four months. This data allowed us to create accurate contact networks for the early stages of the pandemic. 

\subsection{Updating the network} \label{sub:updateNetwork}
As mentioned in Remark \ref{rm:1}, creating a particular graph and its structure for a large population is both time and memory consuming. Instead, our graph evolves over time. In order to keep track of the transmission of the disease, we only need to include the edges between new exposed individuals $E$ and their contacts.

We choose layer 2 contacts for new added nodes as follows. We first create a list of eligible nodes based on a connectivity distribution between counties. For instance, density, demographic information or mobility data could be used for this purpose. For simplicity, we consider a binary symmetric matrix where its entry $(m,n)$ is equal to $1$ if there are connections between counties $m$ and $n$, and $0$ otherwise; see a toy example in Table \ref{table:familyInfo}. This means that the eligible list for a node includes all IDs from counties that have connectivity with its county. We then randomly choose the remaining number of contacts for each node from the eligible list. Further restrictions can be added when defining the eligible list, such as age group, density and distance between counties. Edges on layer 3 are daily updated in a similar way, excluding preassigned contacts.

\subsection{Probability of infection} \label{sub:probInf}
If the probability that node $j$ infects a susceptible node $i$ at a given day is $\beta_{ij}$, then the probability of infection $p_i$ for node $i$ is given by
\begin{equation*}
p_i = 1-\prod_{j\neq i} (1-\beta_{ij})
\end{equation*}
where $j$ includes all the indices of nodes that can infect node $i$ that particular day. The value $\beta_{ij}$ can depend on the layer, epidemiological state, social distancing, mask use and self-care behaviour, among other factors. Transmission of the virus can also include imported cases where infection occurs in an external location. Non-periodic events can also be considered (as massive events or the creation of clusters due to super spreader events) to study their impact and possible scenarios, where different probabilities $p_i$ can be defined for each particular case.

In our implementation, we consider two binary random variables \texttt{useMask} and \texttt{selfCare} that indicate whether or not an encounter between two nodes includes mask use and self-care, respectively. We assume that a fixed fraction of diagnosed cases do not isolate and that hospitalized individuals cannot infect susceptible nodes. We then have different values for $\beta_{ij}$ depending on the interaction between the two nodes and their attributes. We only need to check susceptible nodes that have infectious contacts. Once $p_i$ is determined, a uniformly distributed pseudo-random number $r_i \in (0,1)$ is generated. If $r_i<p_i$, then node $n_i$ is marked as an exposed individual at the current time step. 

\subsection{Transitions between states} \label{sub:transitions}
At each time step, every node has one of the following states: susceptible, exposed, diagnosed, undiagnosed, hospitalized, recovered or dead; see Figure \ref{fig:model1}. The event where a susceptible becomes exposed is given by the probability of infection as discussed in Section \ref{sub:probInf}. Any other event depends on a {\it probability of transition}. Every node has a different probability to move to a feasible next state depending on the period of time at its current epidemiological state. If such event should happen, we then store the new state and day; if not, the state remains unchanged.

We show a toy example of probabilities of transition for the diagnosed class $O$ in Table \ref{table:transitions}. In this case, if a node has one day as $O$, it has a $40\%$ chance that it will be hospitalized. Otherwise, it will remain as $O$ on day 2. On day three, there are three possibilities: remain at $O$ ($50\%$), move to $H$ ($10\%$) or recover $R$ ($40\%$). Note that each row adds up to 1 (one state should always be chosen according to these probabilities), and eventually all nodes that have been exposed should end up as $R$ or $D$. This approach allows us to include variable stay periods on each class if data is available; average stay periods can be considered otherwise.

\begin{table}[h!]
\centering
\begin{tabular}{c|rrrrrrr}
time in $O$ & $S$ & $E$ & $O$ & $U$ & $H$ & $R$ & $D$ \\ \hline
1    &  &  & 0.6 &  & 0.4 &    &     \\
2    &  &  & 1.0   &  &    &    &     \\
3    &  &  & 0.5 &  & 0.1 & 0.4 &   \\
4    &  &  & 0.1 &  & 0.3 & 0.6 &   \\
5    &  &  &    &  &    & 1.0 &  
\end{tabular}
\caption{Toy example for the probability of transitions between states from the observed class $O$; omitted entries are zero. From compartment $O$, it is possible to move to states $H$ or $R$ as shown in Figure \ref{fig:model1}. \label{table:transitions}}
\end{table}

\section{Numerical results} \label{sec:example}
In this section we present results for two toy examples obtained with a parallel code implemented in Matlab and a Lenovo SR650 Server with two 2.20GHz Intel Xeon Plata 4214 processors. We remark that a generalization of this implementation has allowed us to study disease projections and different scenarios in Costa Rica, where the total population is approximately five million people with feasible running times.

\subsection{Fixed parameters} \label{sec:ex1}
We consider four counties with a population of 1300 nodes; its demographic distribution and the connectivity matrix are shown in Table \ref{table:familyInfo}. We consider initially $E_0$ exposed nodes randomly chosen. Parameters used in these simulations appear in Table \ref{table:parameters2}. We assume that $70\%$ of the population wears a face mask, $35\%$ respects self-care and $60\%$ of exposed nodes become diagnosed. Values for the probability of infection are shown in Table \ref{tab:infection} for layer 2; probabilities are halved for layer 3. Finally, the probabilities of transitions between states are shown in Appendix 1.

\begin{table}[htbp]
\centering
\begin{tabular}{r|c|c|c|c|c}
County & HH & ANIHH & AG1 & AG2 & AG3 \\ \hline
1      & 140 & 3.0 & 20 & 50 & 30  \\ 
2      & 100 & 3.5 & 28 & 44 & 28 \\ 
3      & 80 & 4.0 & 20 & 40 & 40  \\ 
4      & 60 & 3.5 & 40 & 40 & 20 \\ 
\end{tabular} 
\quad
\begin{tabular}{r|cccc}
County & 1 & 2 & 3 & 4\\ \hline
1      & 1 & 1 & 1 & 1 \\ 
2      & 1 & 1 & 0 & 1\\ 
3      & 1 & 0 & 1 & 0 \\ 
4      & 1 & 1 & 0 & 1\\ 
\end{tabular} 
\caption{(left) Toy demographic distribution considered in Section \ref{sec:ex1}. We present the number of households (HH), the average number of individuals per household (ANIHH) and age groups (AG) percentage distribution. (right) Connectivity matrix. Non-zero entries indicate possible connections among counties. \label{table:familyInfo}}
\end{table}

\begin{table}[h!]
\centering
\begin{tabular}{l|l}
Parameter & Interval \\ \hline
Contacts in social network & $[5,15]$  \\
Contacts in sporadic network& $[0,15]$\\
Contacts per day, county 1 & $[5,15]$ \\
Contacts per day, county 2 & $[3,13]$\\
Contacts per day, county 3  & $[1,10]$\\
Contacts per day, county 4 & $[1,10]$\\ 
\end{tabular}
\quad
\begin{tabular}{c|c|c}
Face mask & Self care & $\beta$ \\\hline
No  & No  & 0.21\\
No  & Yes & 0.15\\
Yes & No  & 0.08\\
Yes & Yes & 0.05\\
\end{tabular}
\caption{(left) Network parameters in toy example and (right) probability of infection depending on the type of encounter between two nodes for layer 2. Values on layer 3 are halved; see Section \ref{sec:ex1}. \label{table:parameters2} \label{tab:infection}}
\end{table}

\begin{figure}[h!]
\begin{subfigure}{.5\textwidth}
\centering
\includegraphics[width=\linewidth]{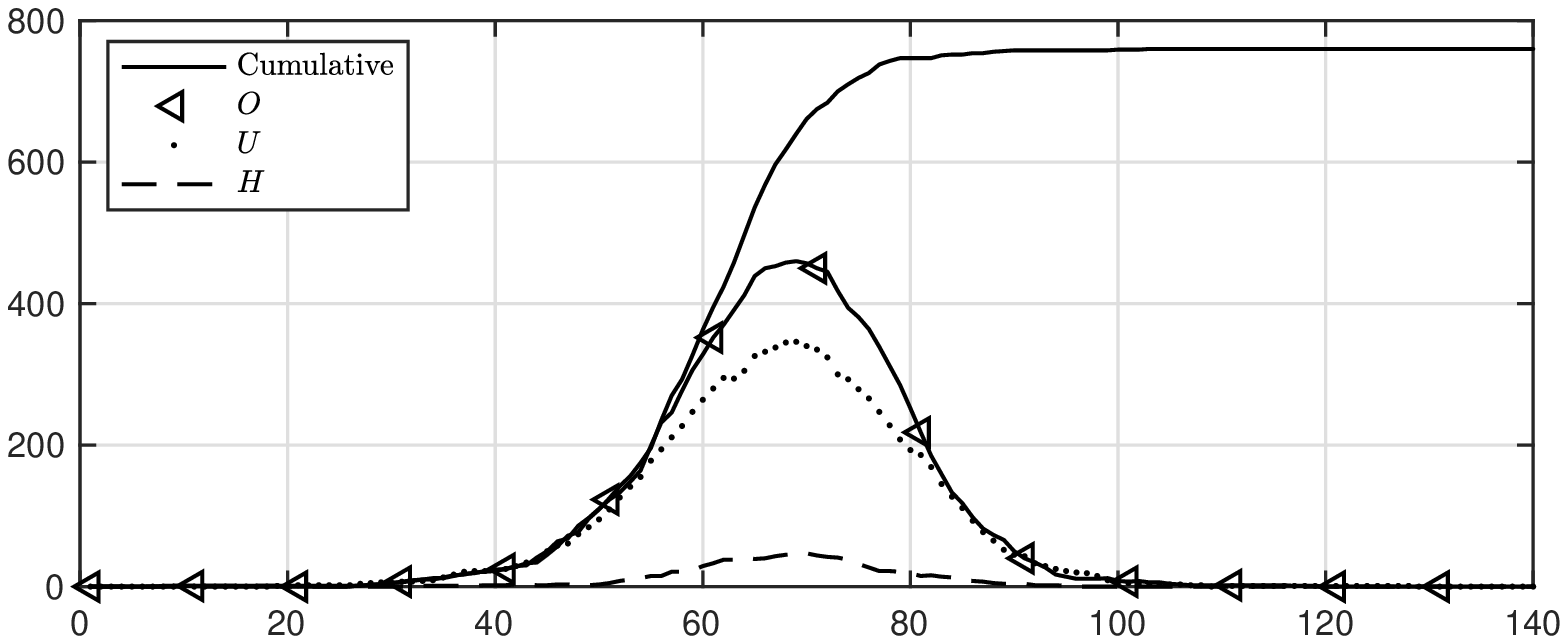}
\captionof{figure}{One simulation}
\label{fig:ex1_1simA}
\end{subfigure}%
\begin{subfigure}{.5\textwidth}
\centering
\includegraphics[width=\linewidth]{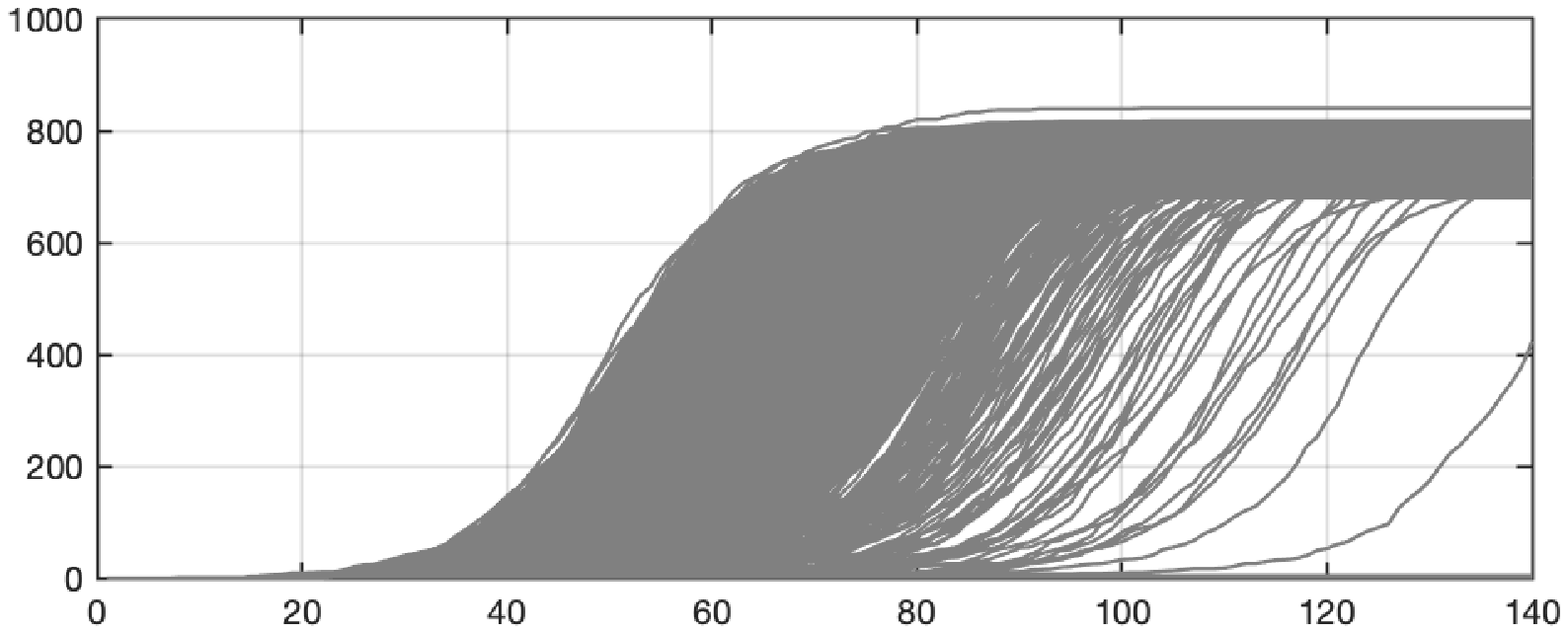}
\captionof{figure}{Cumulative number of cases}
\label{fig:ex1_1simB}
\end{subfigure}
\caption{Experiments with $E_0 = 1$. (left) Cumulative number of cases, $O$, $U$ and $H$ for one simulation. (right) Cumulative number of cases for 1000 simulations; see Section \ref{sec:ex1}. The variability is due to the fact that there is only one initial exposed node. \label{fig:ex1_100sim1E}}
\end{figure}

We present a simulation for $E_0=1$ (only one initial exposed node) in Figure \ref{fig:ex1_1simA}. In Figure \ref{fig:ex1_1simB}, we present 1000 curves for the cumulative number of cases. We remark that in this scenario with only one initial exposed node, there is no outbreak in $7\%$ of the simulations. Meanwhile, in the remaining $93\%$ there are 700 cumulative number of cases on average.

\begin{figure}[t!]
\begin{subfigure}{.5\textwidth}
\centering
\includegraphics[width=\linewidth]{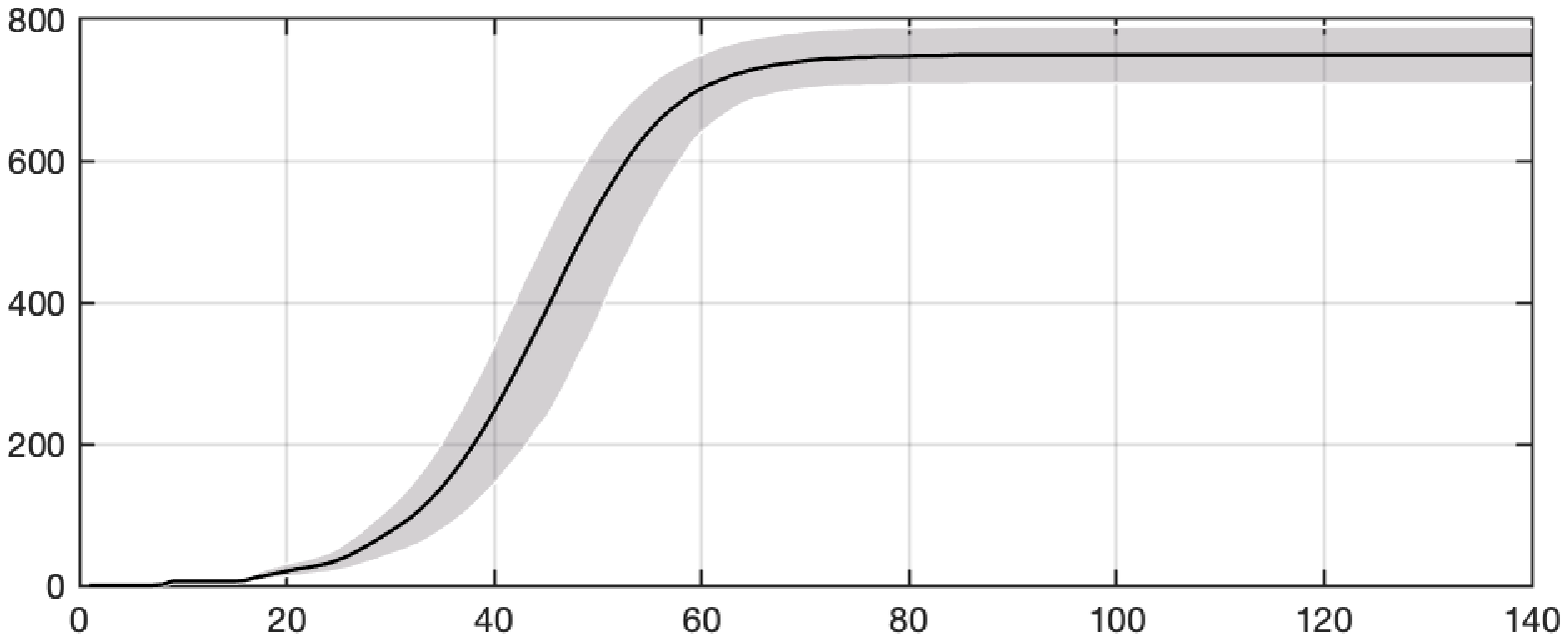}
\captionof{figure}{Cumulative number of cases.}
\label{fig:ex2_1simA}
\end{subfigure}%
\begin{subfigure}{.5\textwidth}
\centering
\includegraphics[width=\linewidth]{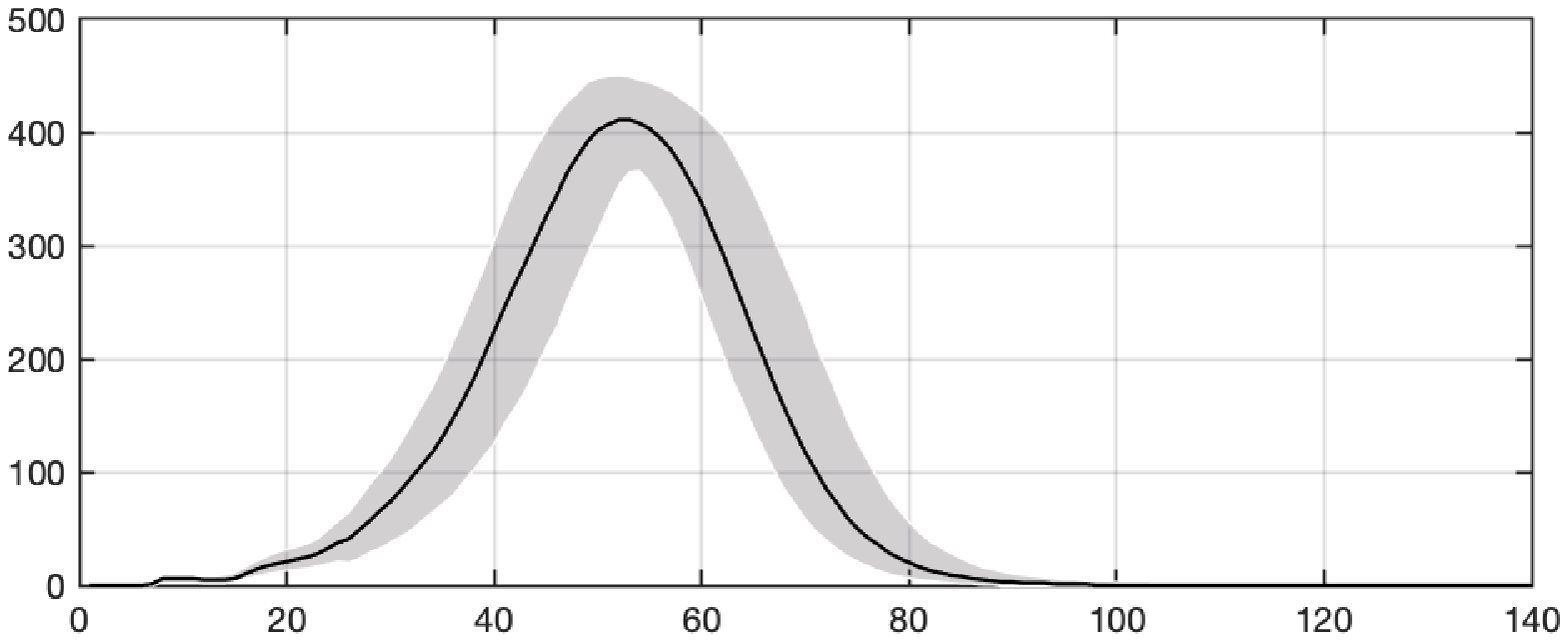}
\captionof{figure}{Observed individuals.}
\label{fig:ex2_1simB}
\end{subfigure}
\begin{subfigure}{.5\textwidth}
\centering
\includegraphics[width=\linewidth]{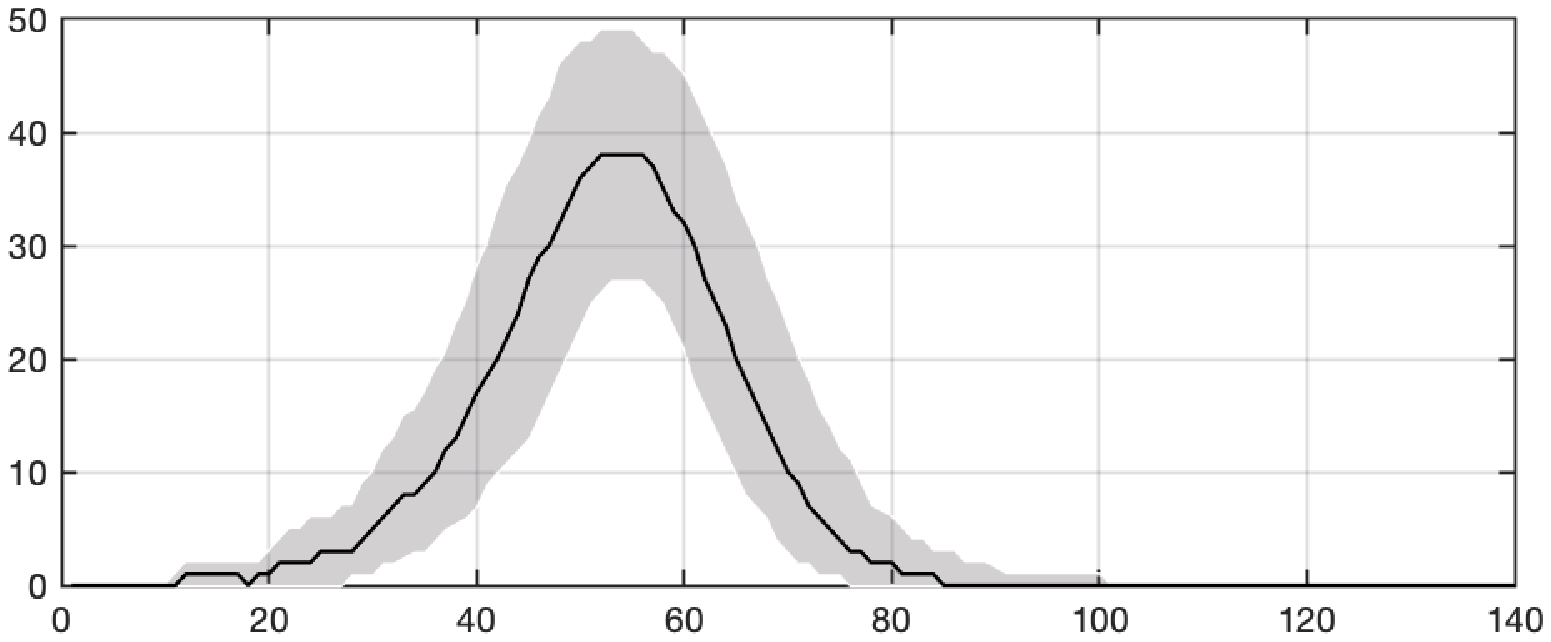}
\captionof{figure}{Hospitalized individuals.}
\label{fig:ex2_1simC}
\end{subfigure}%
\begin{subfigure}{.5\textwidth}
\centering
\includegraphics[width=\linewidth]{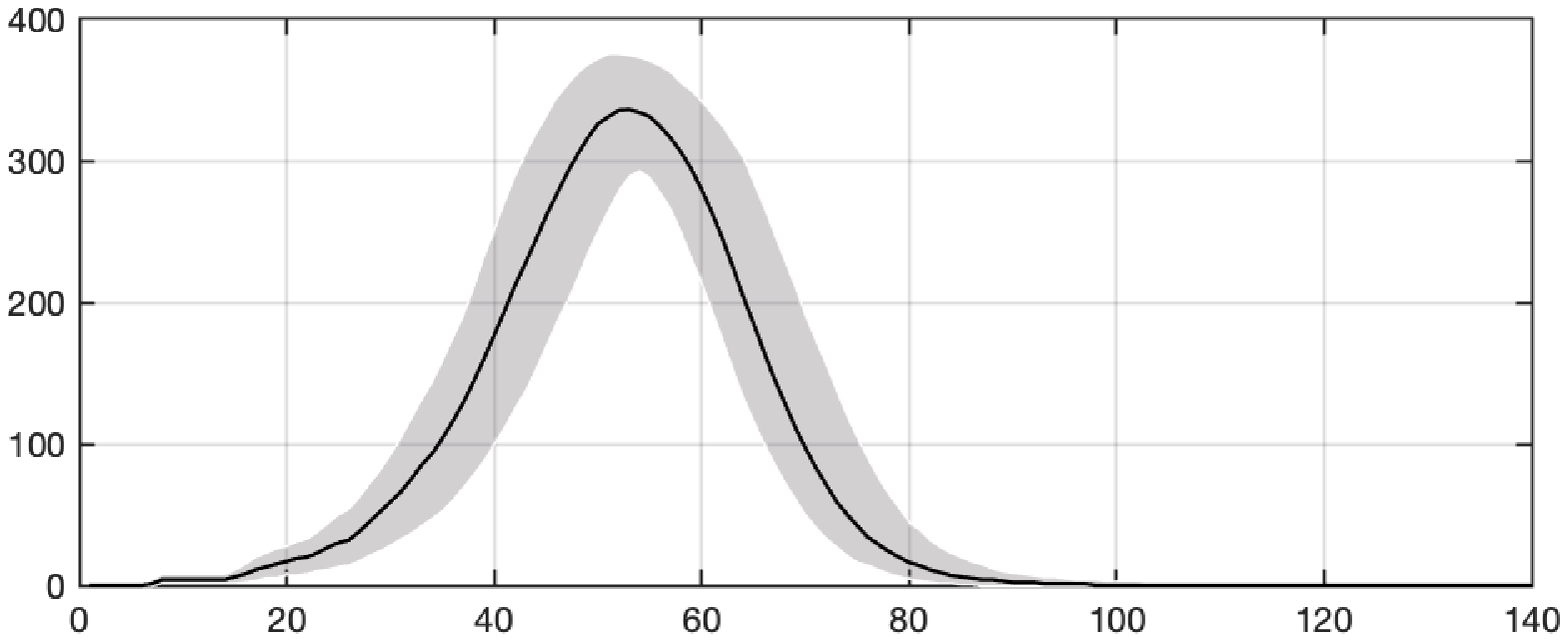}
\captionof{figure}{Undiagnosed individuals.}
\label{fig:ex2_1simD}
\end{subfigure}
\caption{Mean (black line) for 1000 simulations with $E_0=10$. The shaded region corresponds to values between the $5^{\rm th}$ and $95^{\rm th}$ percentiles of 1000 simulations; see Section \ref{sec:ex1}. \label{fig:ex2_100sim10E}}
\end{figure}

For $E_0 = 10$ and the given parameters, there is a one-wave pattern in the numerical results. We present the mean of 1000 simulations with their $5^{\rm th}$ and $95^{\rm th}$ percentiles in Figure \ref{fig:ex2_100sim10E}. Results can also be desegregated by county and/or age group, but we omit these figures for the sake of brevity. They can be obtained by using the app given in \cite{app}.

\subsection{Modeling changes in behaviour} \label{sec:ex2}
\begin{figure}[htb!]
\begin{subfigure}{.5\textwidth}
\centering
\includegraphics[width=\linewidth]{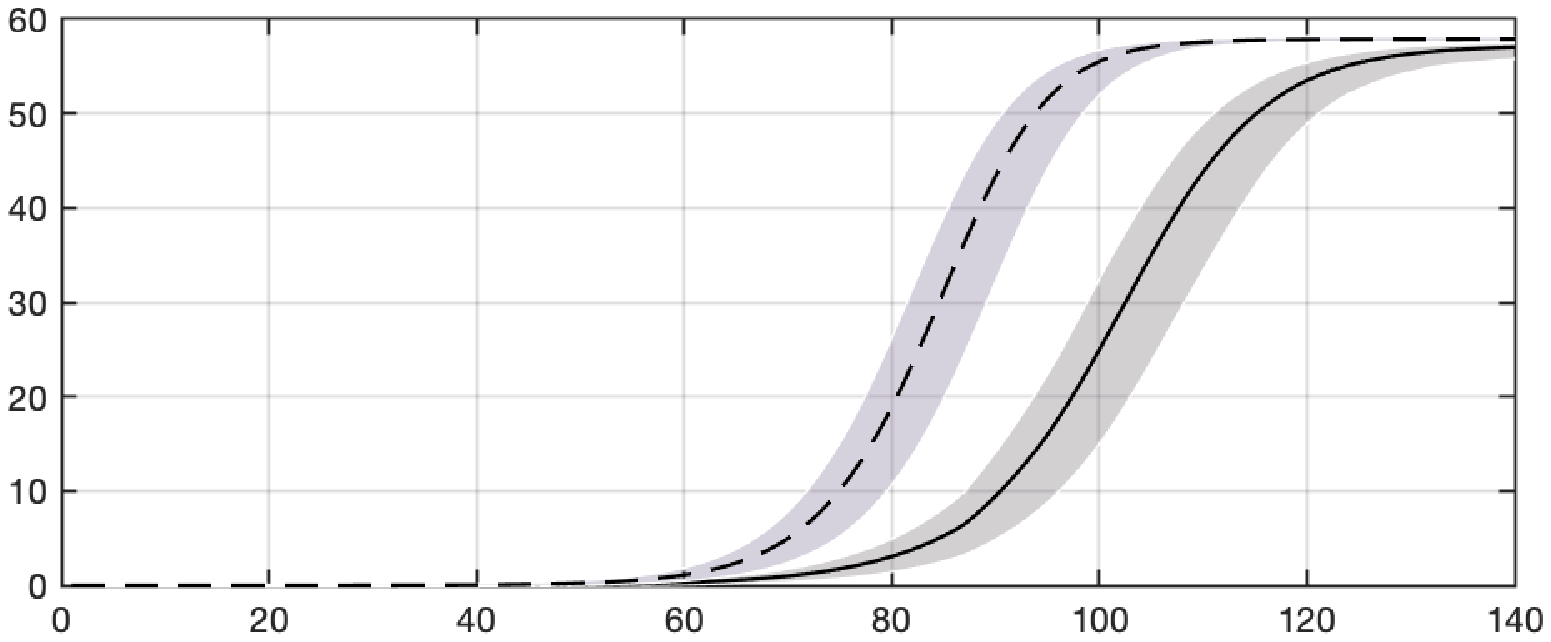}
\captionof{figure}{Cumulative number of cases.}
\label{fig:ex3_1simA}
\end{subfigure}%
\begin{subfigure}{.5\textwidth}
\centering
\includegraphics[width=\linewidth]{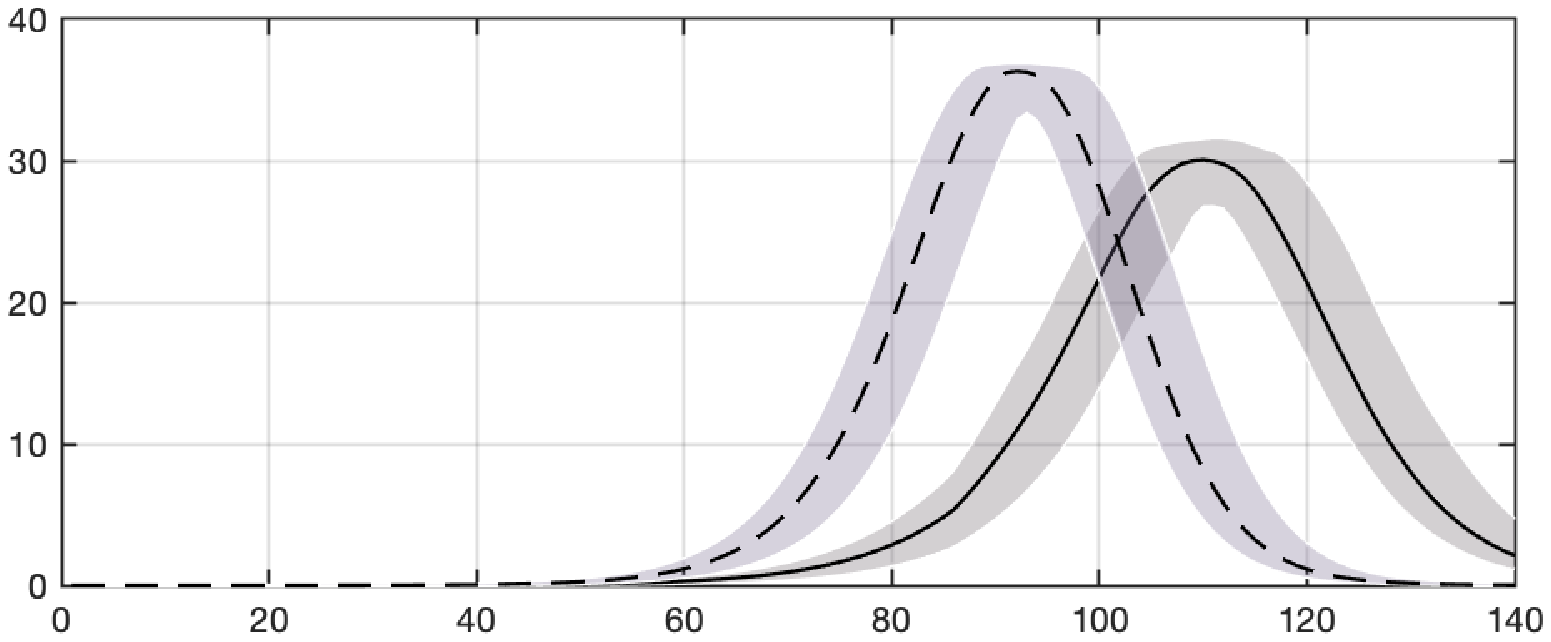}
\captionof{figure}{Observed individuals.}
\label{fig:ex3_1simB}
\end{subfigure}
\begin{subfigure}{.5\textwidth}
\centering
\includegraphics[width=\linewidth]{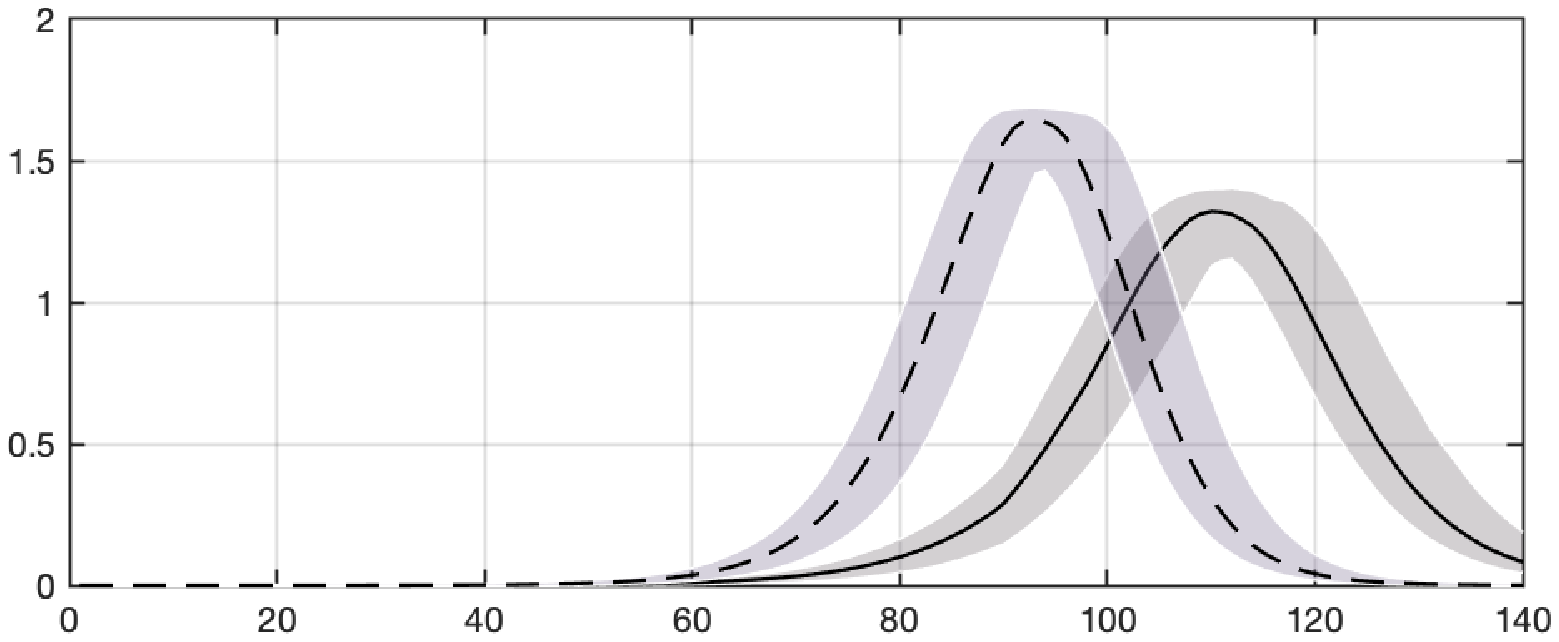}
\captionof{figure}{Hospitalized individuals.}
\label{fig:ex3_1simC}
\end{subfigure}%
\begin{subfigure}{.5\textwidth}
\centering
\includegraphics[width=\linewidth]{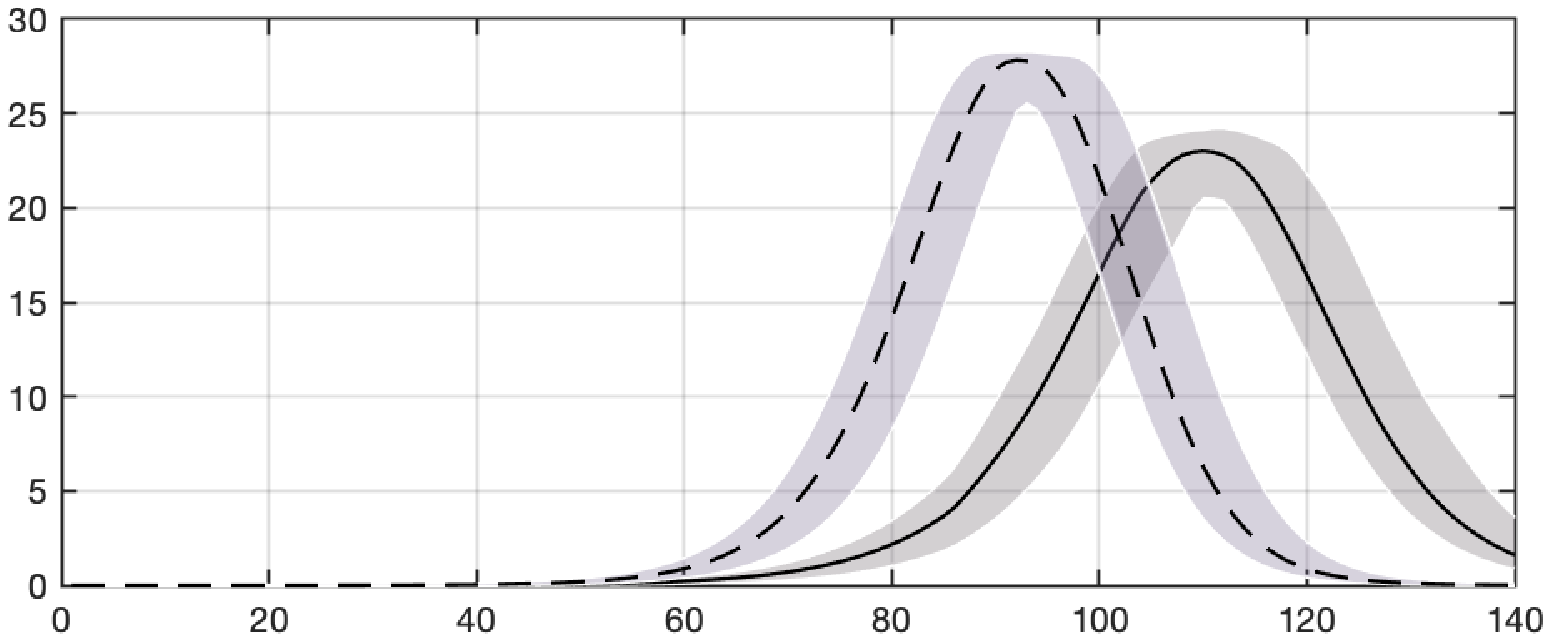}
\captionof{figure}{Undiagnosed individuals.}
\label{fig:ex3_1simD}
\end{subfigure}
\caption{Percentage of cases with no measures (dashed line) and mild restrictions (solid line) for two months starting on day 20. We consider a reduction on the number of daily contacts and an increment in use of masks with $E_0=10$. The shaded region corresponds to values between the $5^{\rm th}$ and $95^{\rm th}$ percentiles of 1000 simulations; see Section \ref{sec:ex2}. \label{fig:ex3_100sim10E}}
\end{figure}

We now consider a population of one million individuals distributed in 15 counties. We impose restrictions on social behaviour by increasing the percentage of individuals that wear a mask and decreasing the number of daily interactions on an period of two months, starting on day 20, emulating the effect of mild regulations. We present results for 1000 simulations in Figure \ref{fig:ex3_100sim10E}, where we compare the effect of such restrictions.

\newpage
\section{Conclusions and final remarks} \label{sec:conc}

We have presented an implementation for a multilayer network to support public health authorities during the Covid-19 pandemic. Capturing the transmission dynamics of the SARS-CoV-2 virus is essential in order to provide information about disease projections and specific data-driven scenarios.

The model exhibits a flexible structure that allowed us to include particular attributes for each node and their contact networks, describing the complexities of social behavior and the specific disease. In this article, the implementation of the general structure of the model is described which can be used in other social and epidemic contexts.

With a health emergency such as the one caused by the SARS-CoV-2 virus, public health authorities were forced to make short-term decisions. Mathematical models help authorities to systematize the transmission mechanisms of the virus and project the expected behavior in the short and medium-term under some assumptions. In particular, a network model is a valuable tool as an input in the decision-making process for health authorities. It has the capability to include social factors and an epidemiological structure, enabling a greater level of detail in the study of the disease. Despite the flexibility of the model, there are limitations that mostly pertain to data availability and its quality.

Implementing and executing these models can be challenging. The availability of information is closely related to the specificity of the question that needs to be answered. Information on the initial conditions for the model and parameter value estimates can be a limiting factor when data is not available or its quality is not optimal. Efficient and fast algorithms are necessary due to the on-demand health scenarios that need to be simulated frequently, and execution times are limited by computational resources.

In an rapidly evolving epidemic as Covid-19, the model must be calibrated frequently, as social behavior is constantly changing and, in many cases, it is unpredictable. The model we have discussed in this paper allows us to straightforwardly adapt and include frequent changes on variables that mimic social behavior. Additionally, the model can be generalized to include other factors, such as the effect of herd immunity and the role of vaccination, as countries take the necessary steps to get back to the new normal.




\section*{Acknowledgments} The authors would like to thank support by the Research Center in Pure and Applied Mathematics and the Department of Mathematics at Universidad de Costa Rica.


\FloatBarrier

\appendix 
\section{Probabilities of transition}
We present the probabilities of transition between epidemiological states used in the toy examples of Section \ref{sec:example}. We remark that there are different probabilities that depend on the age group; see Tables \ref{tab:appE}, \ref{tab:appU}, \ref{tab:appO}, \ref{tab:appH}.

\begin{table}[h!]
\centering
\begin{tabular}{r|c|c|c|c|c|c|c}
Days & 1 & 2 & 3 & 4 & 5 & 6 & 7 \\ \hline
$E$                      & 1.0                   & 1.0                   & 1.0                   & 1.0                   & 1.0                   & 0.8                   &                       \\
$O$                      &                       &                       &                       &                       &                       & 0.1                   & 0.6                   \\
$U$                      &                       &                       &                       &                       &                       & 0.1                   & 0.4                  
\end{tabular}%
\caption{Probabilities from $E$ to $O$ and $U$ depending on the number of days at state $E$; empty entries are zero.}
\label{tab:appE}
\end{table}

\begin{table}[h!]
\centering
\begin{tabular}{l|l|l|l|l|l|l|l|l|l}
Days & 1 & \ldots & 12 & 13 & 14 & 15 & 16 & 17 & 18\\ \hline
$U$  & 1.0    & \ldots & 1.0& 0.9& 0.8& 0.7& 0.5& 0.1&\\
$R$&&&& 0.1& 0.2& 0.3& 0.5& 0.9& 1.0
\end{tabular}
\caption{Probabilities from $U$ to $R$ depending on the number of days at state $E$; empty entries are zero.}
\label{tab:appU}
\end{table}

\begin{table}[h!]
\centering
\begin{tabular}{r|cll|cll|cll}
     & \multicolumn{3}{c|}{Age group 1} & \multicolumn{3}{c|}{Age group 2} & \multicolumn{3}{c}{Age group 3} \\
Days & $O$          & $H$          & $R$     & $O$          & $H$          & $R$     & $O$           & $H$         & $R$     \\ \hline
1    & 1.000          &            &       & 1.000          &            &       & 1.000           &           &       \\
2    & 1.000          &            &       & 1.000          &            &       & 1.000           &           &       \\
3    & 1.000          &            &       & 1.000          &            &       & 1.000           &           &       \\
4    & 0.978      & 0.022      &       & 0.953      & 0.047      &       & 0.629     & 0.371     &       \\
5    & 1.000          &            &       & 1.000          &            &       & 1.000           &           &       \\
\vdots   &   \vdots   &            &       &    \vdots      &            &       &    \vdots    &           &       \\
17   & 1.000          &            &       & 1.000          &            &       & 1.000           &           &       \\
18   &            &            & 1.000     &            &            & 1.000     &             &           & 1.000     \\
\end{tabular}
\caption{Probabilities from $O$ to $H$ and $R$ depending on the number of days at state $E$; empty entries are zero.}
\label{tab:appO}
\end{table}

\begin{table}[th!]
\centering
\begin{tabular}{r|ccc|ccc|ccc}
     & \multicolumn{3}{c|}{Age group 1} & \multicolumn{3}{c|}{Age group 2} & \multicolumn{3}{c}{Age group 3} \\
Days & $H$         & $R$        & $D$        & $H$       & $R$         & $D$         & $H$         & $R$        & $D$        \\ \hline
1    &   1.00        &          &          & 1.00       &           &           & 1.00         &          &          \\
2    &     1.00      &          &          & 1.00       &           &           & 1.00         &          &          \\
3    &     1.00      &          &          & 1.00       &           &           & 1.00         &          &          \\
4    &     1.00      &          &          & 1.00       &           &           & 1.00         &          &          \\
5    &     1.00      &          &          & 1.00       &           &           & 1.00         &          &          \\
6    &     1.00      &          &          & 1.00       &           &           & 0.75      &          & 0.25     \\
7    &     1.00      &          &          & 1.00       &           &           & 0.75      &          & 0.25     \\
8    &     1.00     &          &          & 1.00       &           &           & 0.75      &          & 0.25     \\
9    &    1.00       &          &          & 1.00       &           &           & 1.00         &          &          \\
10   &           &     1.00     &          & 1.00       &           &           & 1.00         &          &          \\
11   &           &     1.00     &          &         & 0.96      & 0.04      & 1.00         &          &          \\
12   &           &     1.00     &          &         & 0.96      & 0.04      & 1.00         &          &          \\
13   &           &     1.00     &          &         & 0.96      & 0.04      & 1.00         &          &          \\
14   &           &     1.00     &          &         & 0.96      & 0.04      &           & 0.75     & 0.25     \\
\end{tabular}
\caption{Probabilities from $H$ to $R$ and $D$ depending on the number of days at state $H$; empty entries are zero.}
\label{tab:appH}
\end{table}

\end{document}